\documentclass[twocolumn]{aastex631}

\usepackage{comment}

\begin{document}

\title{ Identifying Doppelgänger Active Galactic Nuclei across redshifts from spectroscopic surveys}

\correspondingauthor{Swayamtrupta Panda}
\email{swayamtrupta.panda@noirlab.edu}

\author[0009-0003-6558-0317]{Shreya Sareen}\thanks{NOIRLab REU intern}
\affiliation{University of California, Berkeley, 110 Sproul Hall, Berkeley, CA 94720, United States}
\affiliation{Lockheed Martin Rotary Mission Systems, New York, United States}

\author[0000-0002-5854-7426]{Swayamtrupta Panda}\thanks{Gemini Science Fellow}
\affiliation{International Gemini Observatory/NSF NOIRLab,
Casilla 603, La Serena, Chile}



\begin{abstract}

Active Galactic Nuclei (AGNs) are among the most luminous objects in the universe, making them valuable probes for studying galaxy evolution. However, understanding how AGN properties evolve over cosmic time remains a fundamental challenge. This study investigates whether AGNs at low redshift (nearby) can serve as proxies for their high-redshift (distant) counterparts by identifying spectral ``doppelgängers'', AGNs with remarkably similar emission line properties despite being separated by vast cosmic distances. We analyze key spectral features of bona fide AGNs using the Sloan Digital Sky Survey's Data Release 16, including continuum and emission lines: Nitrogen (N V), Carbon (C IV), Magnesium (Mg II), Hydrogen-beta (H$\beta$), and Iron (Fe II - optical and UV) emission lines. We incorporated properties such as equivalent width, velocity dispersion in the form of full width at half maximum (FWHM), and continuum luminosities (1350\AA, 3000\AA, and 5100\AA) closest to these prominent lines. Our initial findings suggest the existence of multiple AGNs with highly similar spectra, hinting at the possibility that local AGNs may indeed share intrinsic properties with high-redshift ones. We showcase here one of the better candidate pairs of AGNs resulting from our analyses. 

\end{abstract}



\section{Introduction} \label{sec:intro}

Active Galactic Nuclei are compact regions at the centers of galaxies where supermassive black holes accrete matter, releasing vast amounts of energy across the electromagnetic spectrum \citep{Antonucci1993, Padovani2017}. Their brightness makes them visible across cosmic distances, allowing astronomers to probe the conditions of the early universe. Despite extensive studies, the evolution of AGN properties—such as black hole accretion rates, gas composition, and feedback mechanisms—remains a subject of ongoing debate. Understanding these changes is crucial for building a comprehensive picture of galaxy formation and the co-evolution of galaxies and their central black holes \citep{2013ARA&A..51..511K, 2024ApJ...977L...9Z}. 

A promising yet underexplored method of studying AGN evolution is through spectral comparison between AGNs at different redshifts. This project focuses on identifying AGN ``doppelgängers'' which are pairs of AGNs that exhibit nearly identical emission line characteristics but exist at vastly different epochs in the universe’s history. In the context of AGNs, a doppelgänger refers to a galaxy or quasar that closely resembles another AGN in observable features, such as line profiles and intensities, while potentially differing in intrinsic physical conditions. If such spectral twins can be found, especially across low and high-redshift populations, they could serve as a bridge for inferring the properties of early-universe AGNs using well-resolved, nearby counterparts.

A typical AGN spectrum contains three primary components: an underlying ionizing continuum from the AGN itself, absorption lines from intervening material, and both broad and narrow emission lines originating from gas in the vicinity of the black hole. The combined information from these spectral features offers critical insights into the dynamics, energetics, and composition of the AGN's immediate environment \citep{2012NewAR..56...93A, 2015ARA&A..53..365N}.

To explore this, we analyze key spectral features—specifically the Mg II, H$\beta$, C IV, and Fe II emission (optical and UV complexes) lines. By comparing parameters like equivalent width, velocity dispersion in the form of full width at half maximum (FWHM), and continuum luminosities (1350\AA, 3000\AA, and 5100\AA) closest to these prominent lines, we aim to identify pairs of AGNs with compelling spectral similarity. The presence of such doppelgängers would suggest a degree of universality in AGN behavior, supporting the use of local AGNs as proxies for those at high redshift and validating the use of local scaling relations \citep{2006ApJ...641..689V, 2013ApJ...767..149B} to infer BH properties at higher redshifts. However, if no such matches are found, it would imply significant evolution in AGN physics, highlighting the need to treat AGNs at different epochs as fundamentally distinct.
\section{Methodology} \label{sec:methodology}

To investigate the possibility of spectral “doppelgängers” among AGNs, this study involves a comparative analysis of AGN spectra across different redshifts using the Sloan Digital Sky Survey (SDSS) Data Release 16\footnote{\url{http://quasar.astro.illinois.edu/paper_data/DR16Q/}} \citep{SDSS16}, which contains 750,414 AGNs. We focus on identifying AGNs that exhibit similar emission-line characteristics despite being separated by vast cosmic distances. The analysis centers on key spectral features critical to understanding AGN physics: Magnesium (Mg II) at 2800Å, Hydrogen (H$\beta$) at 4861Å, Carbon (C IV) at 1549Å, Nitrogen (N V) at 1240\AA, and Iron (Fe II) emission in both UV (2700–2900 Å) and optical (4434–4684 Å) regimes. In addition, we examine continuum luminosities from the accretion disk at 1350 Å, 3000 Å, and 5100 Å. These features are analyzed to extract, for the moment, equivalent widths (which quantify the strength of these emission lines) and full width at half maximum (FWHM) for velocity dispersion. 

\subsection{low- and high-z SDSS sample} \label{sec:sample}
\textit{Are the doppelgangers sky region dependent? Are we finding more doppelgangers in specific regions of the sky? -- }
To investigate potential doppelgängers across cosmic time, the dataset was divided into two redshift-based subsets inspired by a previous work \citep{2022pas..conf...72P} that performed spectral analyses based on an earlier QSO data release \citep{2020ApJS..249...17R}. The low-redshift sample spans a redshift range of 0.617 to 0.89 and includes 57,268 AGNs characterized by emission lines such as Fe II (UV and optical), Mg II, and H$\beta$. The high-redshift sample covers redshifts from 2.09 to 2.25 and consists of 51,927 AGNs, prominently featuring lines like N V, C IV, Fe II (UV), and Mg II. These divisions ensure that comparable spectral features fall within the observable wavelength ranges of each subset (see Figure \ref{fig1}). An important question addressed during this analysis is whether doppelgänger AGNs are region-dependent—that is, whether they tend to occur more frequently in specific areas of the sky. Our findings show that matched AGN pairs are distributed across diverse sky regions, suggesting that their spectral similarities are not a product of localized sky clustering or observational biases, but rather reflect broader, potentially universal physical processes.

\subsection{Doppelganger detection pipeline} \label{sec:piepline}

\begin{figure*}
    \centering
    \includegraphics[width=\linewidth, trim={3cm 0 0 0},clip]{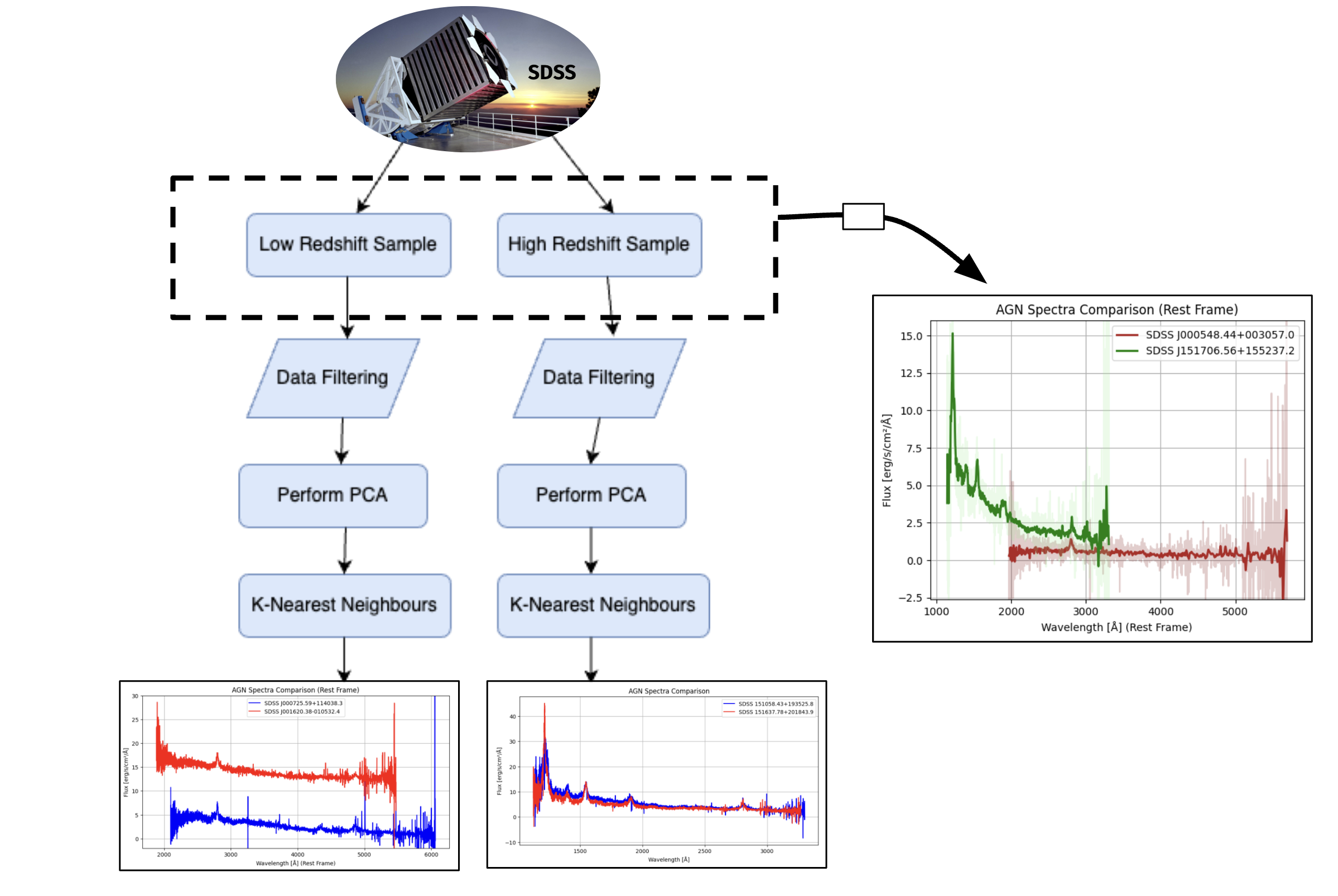}
    \caption{Flowchart for an end-to-end pipeline to retrieve doppelgängers across redshift using SDSS DR16 Spectroscopic Database \citep{SDSS16}. On the bottom left panels, we are highlighting pairs of AGN spectra for the low-redshift sample and the high-redshift sample. On the extreme right, we highlight one pair of AGN spectra resulting from the analysis for the merged sample. We note that the detection of these doppelgängers is independent of the sky region.  }
    \label{fig1}
\end{figure*}

To identify spectral doppelgängers, an end-to-end detection pipeline was developed, as shown with the flowchart in Figure \ref{fig1}. The process begins with data cleaning, including the removal of unreliable or invalid measurements and the selection of relevant spectral features. Following this, dimensionality reduction techniques are applied to condense the high-dimensional feature space while retaining critical information. Principal Component Analysis\footnote{\url{https://scikit-learn.org/stable/modules/generated/sklearn.decomposition.PCA.html}} (PCA, \citealt{Jolliffe2011}) was employed to capture the major sources of variance in the dataset, allowing for a compact and interpretable representation of spectral characteristics. This transformation enabled efficient similarity searches using a nearest-neighbors algorithm based on Euclidean distances in the PCA's hyperparameter space, which systematically compared each AGN to others within the reduced space to identify candidate pairs with the smallest separation. 
This pipeline was applied independently to both low- and high-redshift samples, and later to the merged dataset, allowing for the identification of cross-redshift AGN pairs that share similar spectral profiles.

\section{Results and Discussions} \label{sec:results}
\subsubsection{PCA}

PCA was used to transform the original spectral features into a lower-dimensional space while retaining the maximum variance. For the high-z sample, four principal components captured 100 \% of the spectral variance (PC1 = 40.9 \%, PC2 = 23.1 \%, PC3 = 19.6 \%, PC4 = 16.4 \%), while for the low-z sample, PC1 and PC2 alone accounted for the total variance (PC1 = 76.5 \%, PC2 = 23.5 \%). This enabled 2D visualization and clustering of spectrally similar AGNs.


\subsubsection{Nearest-neighbors algorithm and tweaks}

Once dimensionality was reduced, a nearest-neighbors algorithm was applied to locate AGN pairs with minimal spectral distance in the transformed space. This allowed the identification of $\sim$600 potential doppelgänger pairs. Adjustments to the number of neighbors and distance metrics were made iteratively to balance sensitivity and specificity, ensuring that identified matches had high-quality, interpretable spectral similarities.






When low- and high-z samples were combined, PCA revealed that the first two components (PC1 = 77.1 \%, PC2 = 22.9 \%) were sufficient to represent 100 \% of the spectral variance. Analysis of this merged space identified multiple AGN pairs—each pair consisting of one low-z and one high-z AGN—with strikingly similar spectral features. Emission lines such as Mg II, H$\beta$, Fe II, and C IV aligned well between these matched pairs, including consistency in equivalent widths, FWHM, and line ratios. The similarities, despite differences in redshift and sky position, suggest intrinsic parallels in AGN physics that transcend cosmic time.

\section{Summary and Future Work} \label{sec:summary}

Future work will focus on expanding the scope and depth of the analysis. One key direction is the incorporation of newly released spectroscopic data from the Dark Energy Spectroscopic Instrument (DESI, \citealt{2025arXiv250314745D}), which offers a significantly larger sample of AGNs across a broad redshift range (z = 0–4). 
Additionally, the James Webb Space Telescope (JWST, \citealt{2006SSRv..123..485G}) will be leveraged for its high-resolution infrared capabilities, particularly to observe faint, early-universe AGNs at redshifts beyond z $\gtrsim$ 4. 
Lastly, we aim to refine the doppelgänger detection pipeline using advanced non-linear dimensionality reduction methods and supervised learning techniques to improve the precision and reliability of AGN matching across epochs.

\begin{acknowledgments}
This research was conducted as part of the “Research Experience in Optical-Infrared Astronomy at NOIRLab in Chile” REU Site funded by NSF Award Number 2349023. SP is supported by the International Gemini Observatory, a program of NSF NOIRLab, which is managed by the Association of Universities for Research in Astronomy (AURA) under a cooperative agreement with the U.S. National Science Foundation, on behalf of the Gemini partnership of Argentina, Brazil, Canada, Chile, the Republic of Korea, and the United States of America.
\end{acknowledgments}

\vspace{5mm}
\facilities{SDSS}

\software{numpy \citep{numpy}, matplotlib \citep{matplotlib}, scipy \citep{2020SciPy-NMeth}, astropy \citep{2013A&A...558A..33A,2018AJ....156..123A}}

\bibliography{references}{}
\bibliographystyle{aasjournal}

\end{document}